\documentstyle[mprocl,natbib,epsfig]{article}

\newcommand{\numu}[0]{\nu_{\mu}}
\newcommand{\nuebar}[0]{\overline{\nu}_{e}}
\newcommand{\nue}[0]{\nu_{e}}
\newcommand{\numubar}[0]{\overline{\nu}_{\mu}}
\newcommand{\muebar}{(\numu \nuebar)}
\newcommand{\mubare}{(\numubar \nue)}

\newcommand{\Vcd}{\rm |V_{cd}|}
\newcommand{\Vcs}{\rm |V_{cs}|}
\newcommand{\Vub}{\rm |V_{ub}|}
\newcommand{\Vcb}{\rm |V_{cb}|}
\newcommand{\pt}{\rm p_t}

\newcommand{\beq}{\begin{equation}}
\newcommand{\eeq}{\end{equation}}
\newcommand{\beqs}{\begin{eqnarray}}
\newcommand{\eeqs}{\end{eqnarray}}
\def\beq{\begin{equation}}
\def\eeq{\end{equation}}

\newcommand{\stw}{\mbox{$\sin^2\theta_W$}}

\newcommand{\nubmu}{\overline{\nu_{\mu}}}
\newcommand{\nube}{\overline{\nu_{e}}}

\newcommand{\rmt}{\rm\textstyle}

\begin{document}

\pagestyle{plain}

\title{High Rate Neutrino Detectors for Neutrino Factories
\footnote{Presented at the ICFA/ECFA Workshop
"Neutrino Factories based on Muon
Storage Rings" ($\nu$FACT'99), Lyon, France, 5--9 July, 1999.}
}

\author{B.J. King}
\address{Brookhaven National Laboratory, Building 901A,
P.O. Box 5000, Upton, NY11973\\
email: bking@bnl.gov\\
web page: http://pubweb.bnl.gov/people/bking}

\maketitle

\abstracts{
  Three types of high rate neutrino detectors for neutrino
interaction physics at neutrino factories are discussed. High
performance general-purpose detectors might collect event samples
on the order of a billion events or more. This could greatly improve
on existing analyses of neutrino interactions and also lead to new
and important analysis topics including, for example, precise
determinations of the CKM matrix elements $\Vub$ and $\Vcb$.
The potential of such general purpose detectors is illustrated
with reference to a previously discussed detector~\cite{bjkfnal97}
that is structured around a novel and compact vertexing and tracking
neutrino target comprising a stack of CCD pixel devices.
Design ideas and prospects are also discussed for two
types of specialized detectors: (i) polarized targets filled
with polarized solid protium-deuterium (HD),
for unique and powerful studies of the nucleon's spin structure,
and (ii) Fully active liquid tracking targets with masses of
several tonnes for precise determinations of the weak mixing
angle, $\stw$, from the total cross-section for neutrino-electron
scattering. All three detector types pose severe technical
challenges but their utilization could add significantly to
the physics motivation for neutrino factories.
}

%\begin{keyword}
%NUFACT99 ; detectors; muon colliders; neutrino factories; CKM;
%charm
%\end{keyword}
%
%\end{frontmatter}

\section{Introduction}
\label{sec:intro}
%%%%%%%%%%%%%%%%%%%
%%%%%%%%%%%%%%%%%%%

%[I1] topic is detectors for HR neutrino physics
%v. different to LB physics - are not studying the properties
%of the neutrinos themselves but, instead, their interactions
%with quarks and electrons plus using them as probes of nuclear
%structure.
%
%[I2] Detectors are placed as close 
%production ss
%

  Muon colliders and other proposed high-current muon storage rings
can be collectively referred to as neutrino factories. As well as
the long baseline neutrino oscillation studies that are currently
garnering much of the attention, neutrino factories also have
considerable potential for wide-ranging studies involving the physics
of neutrino interactions. Exciting and unique high rate (HR) neutrino
physics could be performed using detectors
placed as close as is practical to the storage ring in order to maximize
the event rate and to subtend the neutrino beam with the
narrowest possible target. Rather than studying the
properties of the neutrinos themselves, such experiments
would instead investigate their interactions with
the quarks inside nucleons and with electrons.
HR detectors needed for these studies form
the topic of this paper.

  The advantages of neutrino beams from stored
muons over traditional neutrino beams are
in some ways even more notable for HR experiments than
for oscillation studies. In particular, the increased neutrino
flux and the much smaller transverse extent close to production
allows the collection of unprecedented event statistics even
in compact fully-active tracking targets backed by high-rate,
high-performance detectors.

  The small transverse extent of the beam at the HR detectors
derives from the production method in neutrino factories.
Muon decays,
\begin{eqnarray}
\mu^- & \rightarrow & \nu_\mu + \overline{\nu_{\rm e}} + {\rm e}^-,
                                             \nonumber \\
\mu^+ & \rightarrow & \overline{\nu_\mu} + \nu_{\rm e} + {\rm e}^+,
                                                 \label{eq:nuprod}
\end{eqnarray}
in the production straight sections of
the muon storage ring
will produce pencil beams of neutrinos with
unique two-component flavor compositions.
The beams from $\mu^-$ and $\mu^+$ decays will be denoted as
$\muebar$ and $\mubare$ in the rest of this paper.
From relativistic
kinematics, the forward hemisphere
in the muon rest frame is boosted into a narrow cone in the laboratory
frame with a characteristic opening half-angle,
$\theta_\nu$, given in obvious notation by
\begin{equation}
\theta_\nu \simeq \sin \theta_\nu = 1/\gamma_\mu =
\frac{m_\mu c^2}{E_\mu} \simeq \frac{0.106}{E_\mu [{\rm GeV}]}.
                                                   \label{eq:thetanu}
\end{equation}
For example, the neutrino beams from 50 GeV muons will have an opening
half-angle of approximately 2 mrad and a radius of only 20 cm at 100
meters downstream from the center of the production straight section.
(This neglects
corrections due to the non-zero width of the muon beam and the length
of the production straight section.)
As an additional advantage besides the increased beam intensity,
the decay kinematics for equations~\ref{eq:nuprod} are precisely
specified by electroweak theory. This enables precisely modeled
and completely pure two-component neutrino spectra for HR physics
at neutrino factories, which is a substantial advantage
over conventional neutrino beams from pion decays, particularly
for high-statistics precision measurements.

 Analysis topics at these novel
beams and detectors at neutrino factories
might extend well beyond traditional neutrino physics topics and
should complement or improve upon many analyses in diverse areas
of high energy and nuclear physics. Section~\ref{sec:gen}
presents an example of a high performance general purpose detector
whose excellent event reconstruction capabilities should address
almost all such analyses, and gives a brief overview of the
physics benefits of these analyses.
There are two important physics topics that might be much
better conducted using specialized
detectors. Polarized targets for spin physics are discussed
in section~\ref{sec:pol}, and section~\ref{sec:nue} introduces
the options for high mass detectors to study neutrino-electron
scattering.

%[I7] point to table reproduced from JH99
%

\begin{table}[htb!]
\begin{center}
\caption{
Specifications, integrated luminosities and event rates 
for the HR targets discussed in this paper and for
50 GeV (500 GeV) muon storage rings. The
approximation is made that the target is situated 100 m (1 km)
downstream from the center of a straight section that
has $N^{ss}_\mu=10^{20}$
decays of 50 GeV (500 GeV) muons. This corresponds to average
neutrino energies of 32.5 GeV (325 GeV) and to approximately
1 (2) years running for storage ring parameters given previously
in the literature~\cite{Lyon, numcbook}.
%..in the literature. See text for further details.
}
\begin{tabular}{|r|ccc|}
\hline
target purpose & general  & polarized  & $\nu-e$ scatt. \\
\hline
material       & Si CCD's & solid HD   & liquid ${\rm CH_4}$ \\
ave. density   & 0.5 ${\rm g.cm^{-3}}$  & 0.267 ${\rm g.cm^{-3}}$
               & 0.717 ${\rm g.cm^{-3}}$ \\
length         & 2 m      &  0.5 m     &  20 m \\
mass/area, $l$ & 100 ${\rm g.cm^{-2}}$ 
               & 13.4 ${\rm g.cm^{-2}}$
               & 1434 ${\rm g.cm^{-2}}$ \\
radius         & 0.2 m    &  0.2 m     & 0.5 m \\
mass           & 126 kg   &  16.8 kg   & 11.25 tonnes \\ 
$\int L {\rm dt}$
               & $6.0 \times 10^{45}\;{\rm cm^{-2}}$
               & $8.1 \times 10^{44}\;{\rm cm^{-2}}$
               & $8.6 \times 10^{46}\;{\rm cm^{-2}}$ \\
no. DIS events:&&&\\
at 50 GeV      & $1.4 \times 10^9$
               & $1.9 \times 10^8$
               & $2.0 \times 10^{10}$\\
at 500 GeV     & $1.4 \times 10^{10}$
               & $1.9 \times 10^9$
               & $2.0 \times 10^{11}$\\
no. $\nu$-e events:&&&\\
at 50 GeV      & $3.5 \times 10^5$
               & NA
               & $7 \times 10^{6}$\\
at 500 GeV      & $3.5 \times 10^6$
               & NA
               & $7 \times 10^{7}$\\
\hline
\end{tabular}
\label{tab:events}
\end{center}
\end{table}

 Table~\ref{tab:events}, which is reproduced from
reference~\cite{numcbook}, displays parameters for examples
of the three types of detectors discussed in the following sections.
It also gives realistic but very approximate integrated luminosities
and event sample sizes for 2 illustrative neutrino factory energies:
50 GeV and 500 GeV. A muon beam energy of about 50 GeV is a likely
choice for a dedicated muon storage ring~\cite{Lyon}, with default
specifications of $10^{20}$ muon decays per year in the production
straight section. Five hundred GeV muons
correspond to a 1 TeV center-of-mass muon collider such as,
for example, that discussed in reference~\cite{numcbook}.

 The event samples in table~\ref{tab:events} are truly impressive.
It is seen that high performance detectors with
fully-active tracking neutrino targets might collect and precisely
reconstruct data samples with of order billions of neutrino-nucleon DIS
interactions -- more than three orders of magnitude larger than any
of the data samples collected using today's much larger and cruder
neutrino targets. Each of the three detector types in
table~\ref{tab:events} will now be discussed further in the
following sections.

\section{Example design for a neutrino detector to study DIS}
%%%%%%%%%%%%%%%%%%%%%%%%%%%%%%%%%%%%%%%%%%%%%%%%%%%%%%%%%%
\label{sec:gen}

% figure example:
%
\begin{figure}[htbp]
  \begin{center}
        \mbox{\epsfig{file=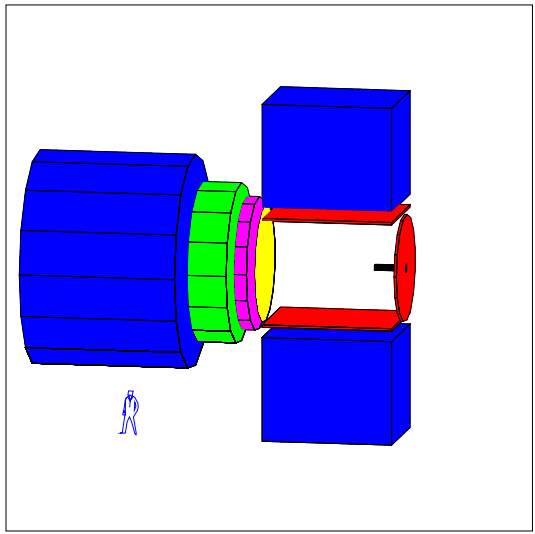,width=7cm}}
  \end{center}
  \caption[]{Example of a general purpose neutrino detector,
reproduced from reference~\cite{bjkfnal97}.
Its scale is illustrated by a human figure in
the lower left corner. The neutrino target is the small
horizontal cylinder at mid-height on the right hand side of the detector. Its
radial extent corresponds roughly to the radial spread of the neutrino pencil
beam, which is incident from the right hand side. The illustration is partially
schematic in that the geometries of the calorimeters and dipole magnet
have been simplified for illustrative purposes.}
  \label{fig:HRdetector}
\end{figure}

 Figure~\ref{fig:HRdetector} shows a general
purpose high rate neutrino detector that might be well matched
to the intense neutrino pencil beams at neutrino factories. This specific
example is reproduced from reference~\cite{bjkfnal97} and it
illustrates the design considerations that might be shared
by other HR detector designs at neutrino factories. A brief overview of
its capabilities will be given in this section; the reader is
referred to the references~\cite{bjkfnal97, numcbook, bjkdetnim}
for more in-depth presentations of its anticipated performance
capabilities.

 As the most striking feature of the detector, the neutrino
pencil beam allows a compact, fully-active precision vertexing
target in place of the kilotonne-scale coarse-sampling
calorimetric targets often used for past and present
high rate neutrino experiments. For example, a 2 meter long
stack of equally-spaced CCD tracking planes with a radius to
match the beam width could contain 1500 planes of 300 micron
thick silicon CCD's, corresponding to a mass per unit area of
approximately 100 ${\rm g.cm^{-2}}$, which is about 5 radiation lengths
or one interaction length. According to table~\ref{tab:events},
even such a modest detector volume might well correspond
to unprecedented neutrino event samples of order a billion,
or even 10 billion, interactions per year.

  The relatively small interaction region of the
CCD target is backed by a hermetic detector that is reminiscent
of many collider detector designs and serves much the same
functions. The enveloping time projection chamber (TPC)
provides track-following, momentum measurements and particle
identification for all charged tracks emanating from the
interactions. Further particle ID might be provided by a mirror
reflecting Cherenkov light to an instrumented back-plane
directly upstream from the
target. Downstream from these, electromagnetic and hadronic
calorimeters use total absorbtion to measure the energies of
individual particles and particle jets and, lastly, iron-core
toroidal magnets will identify muons that have filtered
through the calorimetry.

  Rather than attempting to derive the performance of this
detector for specific physics topics, the rest of this section
will simply present plausibility arguments for its potentially
wide-ranging physics capabilities at neutrino factories, then
quote some more specific conclusions taken from
reference~\cite{numcbook}.

  The dominant interaction processes that provide the physics
content are the charged current (CC) and neutral current (NC)
deep inelastic scattering (DIS) of (anti-) neutrinos off nucleons
($N$, i.e. protons and neutrons)
with the production of several hadrons ($X$):
\begin{eqnarray}
\nu (\overline{\nu}) + N & \rightarrow & \nu (\overline{\nu}) + X
          \;\;\;\;\;\;\;(NC)
                                        \nonumber \\
\nu + N & \rightarrow & l^- + X 
          \;\;\;\;\;\;\;(\nu-CC)       
                                        \nonumber \\
\overline{\nu} + N & \rightarrow & l^+ + X
          \;\;\;\;\;\;\;(\overline{\nu}-CC),
                                        \label{eq:nuint}
\end{eqnarray}
where the charged lepton, $l$, is an electron if the neutrino
is an electron neutrino and a muon for muon neutrinos.
At the many-GeV energies of neutrino factories, these interactions 
are well described as the quasi-elastic
(elastic) scattering of neutrinos off one of the many quarks
(and anti-quarks), q, inside the
nucleon through the exchange of a virtual W (Z) boson:
\begin{eqnarray}
\nu (\overline{\nu}) + q & \rightarrow & \nu (\overline{\nu}) + q
          \;\;\;\;\;\;\;\;\;(NC)
                                        \label{eq:ncnuq} \\
\nu + q^{(-)} & \rightarrow & l^- + q^{(+)} 
          \;\;\;\;\;\;\;(\nu-CC)       
                                        \label{eq:ccnuq} \\
\overline{\nu} + q^{(+)} & \rightarrow & l^+ + q^{(-)}
          \;\;\;\;\;\;\;(\overline{\nu}-CC),
                                        \label{eq:ccnubarq}
\end{eqnarray}
where all quarks, $q$, participate in the NC process but
the CC interactions convert negatively charged
quarks to positive ones for neutrinos and vice versa
for anti-neutrinos, as denoted by
$q^{(-)} \in d,s,b,\overline{u},\overline{c}$ and
$q^{(+)} \in u,c,\overline{d},\overline{s},\overline{b}$.

 It is clear from our experience with collider detectors that
the detector of figure~\ref{fig:HRdetector} could reconstruct
DIS events with at least comparable accuracy and completeness
to, for example, the reconstruction of Z or W decay events at
an e+e- collider detector. The charged leptons from CC interactions
would of course be well measured and, more crucially, the
properties of the struck quark could be inferred from
reconstruction of the hadronic jet it produces. In particular,
the favorable geometry of closely spaced CCD's in the neutrino
target along with their $\sim 3.5\;\mu$m
typical~\cite{SLD} hit resolutions should provide
vertexing of charm and beauty decays that would be
superior to any
current or planned collider detector~\cite{numcbook}.

 The potential richness of neutrino interactions as a
probe of both the nucleon and the weak interaction is
apparent from the 3 experimentally distinguishable processes
of equations~\ref{eq:ncnuq} through~\ref{eq:ccnubarq}, comprising
3 different weightings of the quark flavors probed through
weak interactions involving both the W and Z. Consider, for
comparison, that only a single and complementary weighting
of quarks is probed by the photon exchange interactions of
analogous {\em charged} lepton scattering experiments. 
Past and present neutrino experiments with the more diffuse
neutrino beams from pion decays have suffered from either
insufficient event statistics (e.g. bubble chambers) or
inadequate detector performance (e.g. iron-scintillator
sampling calorimeters) to exploit this rich physics potential.
High rate experiments
at neutrino factories will certainly not lack for statistics --
as evidenced by
table~\ref{tab:events} -- so high performance detectors
should provide the final piece of the puzzle in realizing
the considerable potential of HR neutrino physics.

 Beyond the plausibility arguments given above, more detailed
analyses~\cite{numcbook} suggest the following physics capabilities
for general purpose high rate detectors at neutrino factories:
\begin{itemize}
 \item  the only realistic opportunity, in any physics process,
to determine the detailed quark-by-quark breakdown of the internal
structure of the nucleon
 \item  some of the most precise measurements and tests of
perturbative QCD
 \item  some of the most precise tests of the electroweak theory
through measurements of the electroweak mixing angle, $\stw$,
in neutrino-nucleon deep inelastic scattering, with
uncertainties that might approach a 10 MeV equivalent uncertainty
in the W mass -- i.e. comparable with, and complementary to, the
best measurements predicted for determinations at future colliders
 \item  unique measurements of the elements of the CKM quark mixing
matrix that will
be interesting for lower energy neutrino factories ($\Vcd$ and $\Vcs$)
and will become extremely important ($\Vub$ and $\Vcb$) at
muon beam energies of order 100 GeV and above
 \item  a new realm to search for exotic physics processes
 \item  as a bonus outside neutrino physics, a charm factory
with unique capabilities.
\end{itemize}

%[G2] Other detectors may share design features - small
%target region is analogous to collider geometry and
%this reflects in detector design.
%

%[G3] detector overview:
%- contrast to today's detectors
%- CCD specs
%- summarize rest of detector
%- performance: can handle billions of events/year,
%can now measure hadronic 4-vector to go with Ehad
%(and pmu and theta_mu for CC) - most helpful for NC;
%nearly complete PID - particularly vertex tagging
%of charm (and B) quarks.
%

%[old I6] dominant physics process is DIS off the quarks
%
%[G4] Physics topics (see JH99 & numcbook):
%"Lay bare" the quark structure of nucleons just by
%measuring differential cross-sections
%(and c,b tagging helps also); give SF
%[CKM paragraph] 50 GeV is just threshold for exciting Vub and Vcb
%measurements.
%Many other topics -ref. - best tests of QCD,
%DIS WMA (see other section for target for nu-e measurement),
%new realm for searches, charm factory.
%

%[G5] All these physics topics are intimately coupled with
%the novel high performance detector and beam since
%well-understood beam and excellent event reconstruction
%are required for reduced systematic uncertainties to
%keep pace with the reduced statistical uncertainties.
%

\section{Polarized Nucleon Targets}
%%%%%%%%%%%%%%%%%%%%%%%%%%%%%%%%%%%
\label{sec:pol}

 Neutrinos have intrinsic promise for polarization studies
because they are
100\% longitudinally polarized: neutrinos are always "left-handed"
or "backward" polarized and anti-neutrinos are "right-handed" or
"forward" polarized. Despite this, no past or present neutrino
beam has yet been intense enough or collimated enough for polarized
targets, so polarized neutrino-nucleon DIS appears to have even
more to gain from the improved neutrino beams at neutrino factories than the
non-polarized case presented in the preceding section.

 Until now, the main tool for spin physics studies has been
charged lepton scattering with either polarized electrons or
muons~\cite{Crabb & Meyer}. The capabilities of these experiments
are limited by several factors:
\begin{enumerate}
  \item  the polarization state of the leptons is never 100\%
  \item  the photon exchange interaction provides only a single
probe of the nucleon, as was mentioned in the preceding section
  \item  beam heating of the cryogenic polarized targets places
serious restrictions on their design.
\end{enumerate}

 Very little consideration has yet been given to the design of a
polarized target for neutrino factories or, for that matter, to the design
of the detector that would surround it. The simplest design solution
is to copy the targets used in charged lepton spin
experiments. For example, another contribution to this
workshop~\cite{Kevin_pol} discussed designs based on the butanol
target used with polarized muons by the NMC collaboration. The
problem with such a target is that most of its mass resides
in unpolarized nuclei (carbon, in this case) rather than in
the interesting hydrogen atoms so the effective polarization
of the target is diluted by typically an order of magnitude.
It is hoped~\cite{numcbook} that
the absence of significant target heating from the beam will
allow the use of polarized solid protium-deuterium (HD) targets
such as have been used in
experiments with low intensity neutron or photon~\cite{sphice} beams.
The preparation of such targets is a detailed craft~\cite{Crabb & Meyer}
involving doping the targets with ortho-hydrogen and holding
them for long periods of time at very low temperatures and high
magnetic fields, e.g. 30-40 days at 17 T and 15 mK.
In order to avoid building an entire new detector around the
target, an economical solution would be to
place the polarized target immediately upstream from another
detector, such as the general purpose detector described in
the preceding section.

  The fundamental task of such targets at neutrino factories
will be to probe and
quantify the quark and gluon contributions to the longitudinal
spin component, $S_z^N$, of the nucleon. The overall spin
component for forward polarized nucleons is, of course,
1/2 in fundamental units ($\hbar = 1$) and the potential
component contributions are summarized in the helicity sum
rule:
\begin{equation}
S_z^N = \frac{1}{2} =
\frac{1}{2}(\Delta u + \Delta d + \Delta s) +
L_q + \Delta G + L_G,
  \label{eq:spin_sum}
\end{equation}
where the quark contribution is
$\Delta \Sigma = \Delta u + \Delta d + \Delta s$,
$\Delta G$ is the gluon spin and $L_q$ and $L_G$
are the possible angular momentum contributions
from the quarks and gluons circulating in the nucleon.
(In this notation,
$\Delta q \equiv q^{\uparrow \uparrow} - q^{\uparrow \downarrow}$
is the difference between quarks of type $q$ polarized parallel
to the nucleon spin and those polarized anti-parallel, and
similarly for gluons.) The motivation for measuring the individual
terms in equation~\ref{eq:spin_sum} has strengthened following the
experimental observation~\cite{spin_crisis} in 1989 that
only a small fraction of the nucleon spin is contributed
by the quarks, $\Delta \Sigma \ll 1/2$, which has been
considered counter-intuitive and is often referred to as the
nucleon spin crisis.

  Independent of the details of the polarized target,
the experimental procedure for extracting the $\Delta q$'s
at neutrino factories will be rather analogous to the more familiar
extraction of the quark 4-momentum distributions in
conventional non-polarized targets that was alluded to in the
preceding section. The spin ``structure functions'' $g_1$
and $g_5$ will be extracted from differences in the DIS
CC differential cross-sections for the target spin aligned
with, and then opposite to, the neutrino spin direction.
These structure functions correspond to linear combinations
of the quark spin contributions: the parity conserving
structure function, $g_1$, is the sum of quark and anti-quark
contributions (analogous to the 4-momentum contributions
of quarks to the non-polarized
structure function $F_1$) while the parity violating $g_5$
is the difference of quark and anti-quark contributions
(analogous to the non-polarized structure function $F_3$).

  As was the case in the preceding section, the extraction
of the quark-by-quark contributions from the structure
functions should benefit greatly from the richness of
neutrino interactions, with 8 independent structure
functions to be measured: $g_1$ and $g_5$ from both neutrinos
and antineutrinos and for both protons and neutrons.

  The relative advantage over polarized DIS experiments
with charged leptons is particularly evident for the
parity-violating spin structure functions, $g_5$, since
these can only be measured in CC weak interactions.
The only other future opportunity to measure $g_5$ that
has been widely discussed is the possibility of eventually
polarizing the proton beams in the HERA e-p collider.
 Because of kinematic constraints on reconstructing
events, a polarized HERA would be able to make less precise
measurements~\cite{HERApol} for protons in a complementary
kinematic region that will not be accessible to neutrino factories.
It would not provide measurements for neutrons, of course.

 The above method for extracting the various quark spin
distributions is called ``inclusive'' because it sums over
all hadronic final states. Additionally, neutrino factories should also provide
novel and extended capabilities for ``semi-inclusive'' measurements.
In particular, the semi-muonic tagging of charm production is
sensitive to the spin contribution from the strange quarks
in the nucleon, $\Delta s$. In some kinematic regions it may
also provide sensitivity to the spin contribution of the gluon,
$\Delta G$. Such a capability, if realized, would be very valuable
in helping to solve the spin crisis since $\Delta G$ is extremely
difficult to measure and yet it is the leading suspect for
providing the bulk of the nucleon's spin.

\section{Neutrino detector for neutrino-electron elastic scattering}
%%%%%%%%%%%%%%%%%%%%%%%%%%%%%%%%%%%%%%%%%%%%%%%%%%%%%%%%%%%%%%%%%%%%%%
\label{sec:nue}

%[E1]  The other physics topic that would benefit from a specialized
%detector is the precise determination of the WMA from the
%measurement of the cross-section for neutrino-electron scattering.
%[formula].
%The experimental signature for the process is a single electron
%track with very low pt with respect to the beam direction,
%[formula], and no other activity in the detector.
%
%

  The other physics topic that would benefit from a specialized
detector is the precise determination of the weak mixing angle,
$\stw$, from the measurement of the cross-section for neutrino-electron
scattering:
\begin{equation}
\nu e^- \rightarrow \nu e^-.
\label{eq:nue}
\end{equation}
This is an interaction between point elementary particles
with a precise theoretical prediction for its cross section
as a function of $\stw$ so statistical and experimental
uncertainties will always dominate over the theoretical
uncertainty.

 Determination of the absolute cross section for the
process of equation~\ref{eq:nue} provides a less traditional
way of measuring $\stw$ with neutrinos
than the neutrino-nucleon DIS scattering method mentioned in
section~\ref{sec:gen} and is complementary since the two
measurements have different sensitivities to exotic physics
processes. The best measurement so far from neutrino-electron
scattering was performed with a finely segmented sampling
calorimetric target by the CHARM~II collaboration at CERN, with the
result\cite{CHARM2}:
\begin{equation}
\stw=0.2324\pm0.0058{\rmt (stat)}\pm0.0059{\rmt (syst)}.
\end{equation}
The systematic component of the 3.6\% total uncertainty
is due mainly to beam normalization and background
uncertainties. Background uncertainties at this level
may well be intrinsic to sampling calorimetric targets
so the new approach of a tracking detector is probably
required to obtain much more precise measurements at
neutrino factories.

  The experimental signature for the process in a tracking
detector is a single electron
track with a very low transverse momentum with respect to
the beam direction, $\pt < \sqrt{2 m_eE_\nu}$, and no other
activity in the detector.
 The two experimental challenges that motivate a dedicated
detector are:
\begin{enumerate}
  \item  the cross section for neutrino interactions with electrons
is three orders of magnitude below the dominating process of DIS
interactions with nucleons. Even at neutrino factories this would
require a relatively massive detector -- perhaps several
tonnes -- to obtain sufficiently large event samples.
  \item  the crucial measurement of the electron $\pt$'s must
be obtained before the electron initiates an electromagnetic
shower and this distance scale is characterized by the
radiation length of the tracking medium. This effectively
restricts the target to elements with particularly
low atomic numbers (Z) because the radiation length
scales inversely as ${\rmt Z^2}$.
\end{enumerate}
Other desirable characteristics for the detector are
a fully-active tracking medium with good position
resolution, a magnetic field to verify the negative
charge of the electrons and a fast read-out to
minimize pile-up from the DIS background events.

 An attractive target/detector option is a cylindrical tank
containing a low-Z liquid that can form tracks of ionization
electrons and drift them to an electronic read-out. The
choice of the liquid requires a more detailed survey of
low-Z liquid properties than has so far been conducted.
Working up from the lowest Z, liquid hydrogen (Z=1) is
unfortunately ruled out because of insufficient electron
mobility. Liquid helium (Z=2) also suffers from poor mobility
and potentially difficult operation because it lacks the
ability to self-quench.

  Liquid methane and other saturated alkanes appear to be
good candidates for the tracking medium as they contain only carbon
(Z=6) and hydrogen (Z=1) and sufficiently pure samples are capable
of transporting electrons over large distances. Experimental
studies~\cite{aprile} of electron transport in methane have been
successful enough to suggest its use in TPC detectors of up to
several kilotons. It is liquid at atmospheric pressure between
--182.5 and --161.5 degrees centigrade and has a density of
0.717 g/cm$^3$ and a radiation length of 65 cm. Heavier alkanes
that are liquid at room temperature, such as octane, would be
superior for safety and convenience if they can be maintained
at sufficient purity for good electron transport, and this
deserves further study. Finally, liquid argon (Z=18) also deserves
further consideration~\cite{Williscomm} despite having a radiation
length of only 14 cm since its suitability as a large-volume
tracking medium has been convincingly demonstrated through
prototyping for the multi-kilotonne ICANOE~\cite{ICANOE}
neutrino oscillation detector.

\begin{figure}[htbp]
  \begin{center}
        \mbox{\epsfig{file=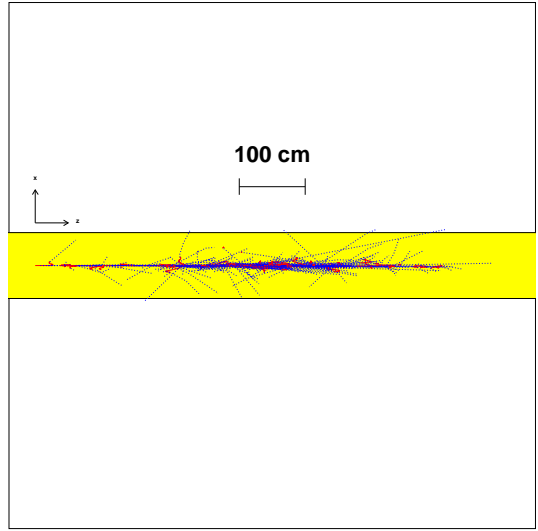,width=7cm}}
  \end{center}
  \caption[]{Monte-carlo generated picture of
a neutrino-electron scattering interaction in
liquid methane. The view
is perpendicular to both the beam direction and the magnetic field.
The red tracks are electrons or positrons. The blue tracks are photons,
and these won't actually be seen in the detector.
}
  \label{fig:nue_in_methane_xz}
\end{figure}

\begin{figure}[htbp]
  \begin{center}
        \mbox{\epsfig{file=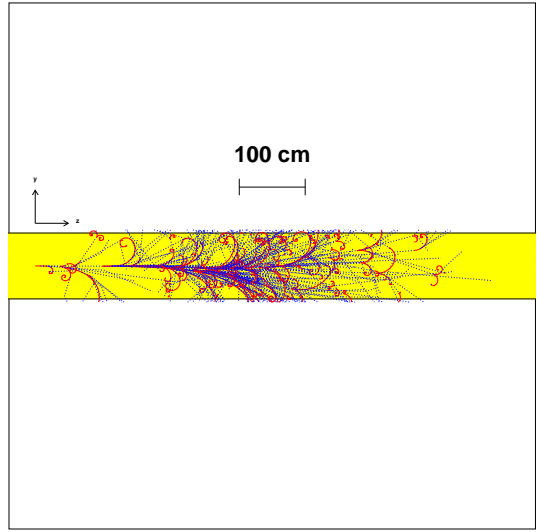,width=7cm}}
  \end{center}
  \caption[]{Monte-carlo generated picture of
a neutrino-electron scattering interaction in
liquid methane. Same event as the
immediately preceding
figure except that the view
is now parallel to the magnetic field.}
  \label{fig:nue_in_methane_yz}
\end{figure}

 Example monte carlo-generated event pictures for neutrino-electron
scattering events in liquid methane are shown in
figures~\ref{fig:nue_in_methane_xz} and~\ref{fig:nue_in_methane_yz}.
Essentially all of the $\pt$ and charge sign
information on the electron is contained
in the initial track at the upstream (left hand side) of the display.
% figure example:
%

% figure example:
%

  A time projection chamber (TPC), as used in
ICANOE, is the best established readout option but may run
into problems with event pile-up due to the large drift
distances characteristic of this readout geometry. A faster read-out
alternative that has been suggested by Rehak~\cite{Rehak}
uses printed-circuit kaptan strips to provide more channels
and, hence, shorter drift distances.

 The two other big experimental challenges for the measurement
besides event pile-up are background rejection
and benchmarking the signal event rate to precisely
predictable flux normalization processes, which will
now be discussed in turn.

  The dominant DIS background events will usually be readily
distinguishable from the signal due to their high track multiplicity
at the primary vertex. Instead, the most difficult backgrounds
will come from low-multiplicity neutrino-nucleon scattering
events such as quasi-elastic neutrino-nucleon scattering:
\begin{equation}
{\rm \nu N \to  l^\pm N'},
\label{eq:quasielastic}
\end{equation}
where $N'$ is an excited state of the nucleon $N$.
A tracking detector with
very good $p_t$ resolution is needed to resolve the signal peak
from the much broader background distributions from such events.

 Flux normalization~\cite{numcbook} should be less difficult for
the $\muebar$ beam due to the availability of
two theoretically predictable normalization processes
involving muon production off electrons:
\begin{eqnarray}
\numu e^-\to\nue \mu^- \\
\nube e^-\to\nubmu\mu^-.
\label{eq:etomu}
\end{eqnarray}
The $\mubare$ beam is more problematic, probably requiring an additional
stage of relative flux normalization back to the $\muebar$ flux
using the relative sizes of the event samples for the quasi-elastic
neutrino-nucleon scattering process of equation~\ref{eq:quasielastic}.
This requires the detector to also measure very low-$\pt$ muons from
the processes of equation~\ref{eq:etomu} and~\ref{eq:quasielastic}
which should in practice be less difficult than the signal process
for the detectors under consideration.

% It therefore provides measurements of
%$\stw$ that will be essentially
%limited only by statistics (3 orders of magnitude down from DIS)
%and by ingenuity in minimizing the experimental uncertainties.

 Table~\ref{tab:events} gives signal event sample sizes in the range of
millions to tens-of-millions of events for an 11-tonne liquid methane
detector. This corresponds~\cite{numcbook} to the impressive limiting
statistical uncertainties of $\Delta \stw = 0.000\,3$ and $0.000\,1$
for the $\muebar$ and $\mubare$ beams, respectively, at the 50 GeV
neutrino factory,
and to $\Delta \stw = 0.000\,1$ and $0.000\,03$ for the 500 GeV
neutrino factory. With negligible theoretical uncertainties for this process
the experimental challenge in approaching these statistical limits
rests largely on the design of a specialized detector that can
minimize the experimental uncertainties.

%  All-in-all, the design of a dedicated detector for
%neutrino-electron scattering appears to be very challenging
%but probably feasible.

\section{Conclusions}
\label{sec:concl}
%%%%%%%%%%%%%%%%%%%
%%%%%%%%%%%%%%%%%%%

  The prospects for short-baseline high rate neutrino physics
at future neutrino factories is substantial and is tightly coupled
to the development of novel high performance neutrino detectors
that exploit the uniquely intense and collimated neutrino beams
at these facilities.

  Three types of high rate neutrino detectors have been discussed
in this paper:
\begin{enumerate}
  \item  general purpose detectors featuring, for example, a fully
active CCD vertexing and tracking target and with a backing detector
of similar complexity and performance to some collider detectors.
Such detectors would have wide-ranging potential for extending
neutrino physics well beyond its traditional bounds.
  \item  polarized targets that might map out the quark-by-quark
spin structure of the nucleon and, perhaps, also determine the gluon
contribution to the nucleon's spin. Cryogenic targets
of solid hydrogen might have much superior performance to the
conventional polarized targets used in charged lepton scattering
if their considerable design challenges can be negotiated.
  \item  fully active tracking targets comprising several tonnes
of low atomic number liquids hold promise for one of the most
precise tests of the electroweak interaction, through the
determination of the weak mixing angle, $\stw$, from the total
cross-section for neutrino-electron scattering.
\end{enumerate}

 The designs for all three detector types are both exciting and
very challenging. Designs for general purpose detectors have been
presented only at the conceptual level and those for the two
specialized target types have not even proceeded that far. Given
the levels of complexity and challenge, there is both an opportunity
and a need to soon begin the design work towards realizing these
detector options at the first neutrino factory facility.
The expected experimental conditions at neutrino factories are so novel
and impressive that there are sure to be many surprises along
the way.

\section{Acknowledgments}

 The author has benefitted from many discussions with his
co-authors on reference~\cite{numcbook}. A discussion on
spin physics with M. Velasco was also valuable. The organizers
and secretariat of NUFACT99 are to be commended for a
well-organized and stimulating workshop.

This work was performed under the auspices of
the U.S. Department of Energy under contract no. DE-AC02-98CH10886.

%%%%%%%%%%%%%%%%%%%%%%%%%%%%%%%%%%%%%%%%%%%%
%%                                         %
%%       APPENDIX PART (IF NEEDED)         %
%%       (UNCOMMENT THE NEXT LINES)        %
%%%%%%%%%%%%%%%%%%%%%%%%%%%%%%%%%%%%%%%%%%%%

%% \appendix
%% \section{APPENDIX}
%% Write here any appendix equations or text.


\begin{thebibliography}{999}

\bibitem{numcbook} I.I. Bigi {\it et al.},
``The potential for High Rate Neutrino Physics at Muon
Colliders and Other Muon Storage Rings'', in preparation
for publication in Physics Reports.
\bibitem{Lyon} These proceedings.
\bibitem{bjkfnal97}
   B.J. King,
   ``Neutrino Physics at a Muon Collider'',
   {\em Proc. Workshop on Physics at the First Muon Collider
   and Front End of a Muon Collider}, Fermilab, November 6-9, 1997.
\bibitem{bjkdetnim} B.J. King, paper in preparation.
\bibitem{SLD}  K. Abe {\em et al.}, ``Design and performance
of the SLD vertex detector: a 307 Mpixel tracking system'',
{\em NIM} {\bf A400} (1997), 287-343.
\bibitem{Crabb & Meyer} D.G. Crabb and W. Meyer,
 ``Solid Polarized Targets for Nuclear and Particle Physics
Experiments'', {\em Ann. Rev. Nucl. Part. Sci.} {\bf 47} (1997), 67-109.
\bibitem{Kevin_pol}  K.S. McFarland, ``Nucleon structure, weak
    interactions, neutrino properties: where do we stand
    and where do we want to go?'', these proceedings.
\bibitem{sphice} D. Babusci {\it et al.},
 {\em Proc. 11th Int. Symp. High-Energy Spin Phys., AIP Conf.
Proc.} {\bf 343} (1995), 523.
\bibitem{spin_crisis} J. Ashman {\it et al.}, the EMC Collaboration,
 {\em Phys. Lett. B} {\bf 206} (1988), 364.
\bibitem{HERApol} See, for example, A. Deshpande,
``The Physics Case for Polarized Protons at HERA'', hep-ex/9908051.
\bibitem{CHARM2} P.~Villain {\em et al.\,}, {\em Phys. Lett.}
{\bf B335} (1994), 246.
\bibitem{aprile} Aprile, Giboni and Rubbia {\em NIM} {\bf A253} (1987), 273.
\bibitem{Williscomm} Private Communication with W.J. Willis.
\bibitem{ICANOE} The ICANOE Collaboration, home page
http://pcnometh4.cern.ch/ .
\bibitem{Rehak} Private communication with P. Rehak.

\end{thebibliography}
\end{document}